\documentstyle[12pt]{article}
\begin{document}

\def\simgt{\ \raisebox{-.25ex}{$\stackrel{>}{\scriptstyle \sim}$}\ }
\def\simlt{\ \raisebox{-.25ex}{$\stackrel{<}{\scriptstyle \sim}$}\ }
\def\anti#1{\overline#1}
\def\solar{_\odot}

\begin{titlepage}
\vspace{.3in}
\begin{center}
{\Large{\bf The anthropic principle and the mass scale of the Standard
Model }}\\
\vskip 0.3in
{\bf V. Agrawal${}^a$, S.M. Barr${}^a$, John F. Donoghue${}^b$
 and D. Seckel${}^a$}\\[.2in]
a){\it Bartol Research Institute \\
University of Delaware, Newark, DE 19716\\}
b){\it Department of Physics and Astronomy, \\ 
University of Massachusetts,
Amherst, MA 01003 \\}

\end{center}
\vskip 0.4in

\begin{abstract}
In theories in which different regions of the universe can have different 
values of the the physical parameters, we would naturally find ourselves 
in a region which has parameters favorable for life. We explore the range 
of anthropically allowed values of the mass parameter in the Higgs 
potential, $\mu^2$. For $\mu^2<0$, the requirement that complex elements 
be formed suggests that the Higgs vacuum expectation value $v$ must have 
a magnitude less than 5 times its observed value, For $\mu^2>0$, baryon 
stability requires that $|\mu|<<M_P$, the Planck Mass. Smaller values of 
$\left| \mu^2 \right|$ may or may not be allowed depending on issues of element 
synthesis and stellar evolution. We conclude that the observed value of 
$\mu^2$ is reasonably typical of the anthropically allowed range, and 
that anthropic arguments provide a plausible explanation for the closeness 
of the QCD scale and the weak scale.   

\end{abstract}
\end{titlepage}

\section{Introduction}

Some of the major puzzles of particle physics
and cosmology concern parameters which are much smaller
than their expected ``natural" size [1]. For example,
the Yukawa coupling constant of the electron, $\lambda_e$,
is about $2 \times 10^{-6}$.
The QCD ``vacuum angle", $\overline{\theta}$, is
experimentally bounded to be less than about 
$10^{-9}$. Perhaps the mother of all such
small-number puzzles is the cosmological constant
problem [2]. The cosmological constant, $\Lambda$, in
natural gravitational units ({\it i.e.} in units
of $M_P^4$, where $M_P \sim 10^{19}$ GeV is the ``Planck mass") is
known to be less than about $10^{-120}$. 
All of these dimensionless ratios would naturally
be expected to be of order unity in the absence
of some dynamical mechanism or symmetry principle
that determined them to be small.

For some of these small numbers fairly simple
conventional particle physics explanations are
available. In particular, viable models exist
for explaining the smallness of $\overline{\theta}$
and $\lambda_e$. Other small
numbers are harder to explain conventionally.
For example, although one can find ways (involving supersymmetry)
to explain why the cosmological constant is as small as 
(1 TeV)$^4$ (or $10^{-64}$ in natural units), that is still
$56$ orders of magnitude larger than the observational bound.

In recent work, Weinberg has addressed the question
of whether the Anthropic Principle can explain the
smallness of the cosmological constant [3]. Roughly
stated, the Anthropic Principle [4] says that the parameters
of the universe that we observe are governed by the
requirement that they must be able to support intelligent 
life, as otherwise we would not exist to observe our 
universe. If there is only one single ``universe", with
the same laws and parameters everywhere, this is far from
a satisfactory physical explanation. However, it has been
realized that some physical theories can support the
existence of separated domains in the universe in which
different parameters and even different gauge groups
are applicable. For example, in chaotic inflation [5] 
different domains have different Higgs vacuum expectation
values, selecting different effective particle physics
theories. Such domains could be regarded as, effectively,
different universes. This idea that multiple ``universes" can exist 
takes the Copernican revolution to the ultimate limit --- 
even our universe may not be unique. In a 
multiple universe theory, the anthropic requirement that we
live in a universe with viable parameters is as natural
as is the good fortune that we happen to live on a 
planet that has a temperature ideal for life.

Discussions of the Anthropic Principle often end up
dangerously close to being non-scientific. However,
Weinberg shows how calculations can be done based
on mild forms of the the Anthropic Principle --- or
perhaps more accurately the multiple universe hypothesis ---
which allow one to assess whether they are able to explain
the smallness of a parameter such as the cosmological
constant [3] Weinberg examines the requirement that the 
evolution of the universe be such that matter clumps into galaxies,
and shows that this only occurs for a range of values of the 
cosmological constant. He then calculates the mean value of
viable $\Lambda$'s. If the actual value of $\Lambda$
in our universe turns out to be very much smaller than 
this mean value, one would conclude that this form of the 
Anthropic Principle does {\it not} provide sufficient 
explanation for the magnitude of $\Lambda$. Unfortunately,
the results are not yet conclusive, as neither the
mean viable value nor the experimental value of $\Lambda$ is yet
well determined.  

The greater part of this paper will be devoted to applying
anthropic arguments to a single parameter of the Standard
Model of particle physics, namely $\mu^2$, the mass parameter of the Higgs
potential. Like the cosmological constant, this parameter is many 
orders of magnitude smaller than its ``natural scale" and has
so far no completely satisfactory conventional particle physics
explanation. Its smallness (about $10^{-34}$ in Planck units) is
considered one of the major puzzles in particle physics, and is
often called the ``fine tuning problem''
or (in the context of grand unified theories)
the ``gauge hierarchy problem" [6].
We will assume that the Standard
Model is the correct theory of particle interactions in the limit
that effects from the Planck scale or the unification scale
are neglected. In the next several sections of the paper we will see
what the consequences are of varying a single parameter, $\mu^2$, 
of that model [7].
In particular we will ask whether for different
ranges of $\mu^2$ complex elements, which are presumably required for
the emergence of life, can (1) exist, and (2) be actually formed
in significant quantities in the evolution of the universe.

It is important that we do not attempt to modify the Standard Model
by adding new physics at low energies. First, 
to do so would make the space of possibilities too large to make 
meaningful anthropic arguments. And, second, the need for new low-energy
physics would be unclear if the magnitude of $\mu^2$ could be explained 
without it.

The reasons for focusing particularly on $\mu^2$ will be explained more 
fully below, but it is also interesting to look at certain other parameters
of the Standard Model from an anthropic perspective, and this we do in
section 6, where we examine the ratios of the quark masses.

\section{The Higgs mass parameter $\mu^2$}

Of all the parameters of the Standard Model, $\mu^2$
stands out in a number of ways. First, it is the only one which is 
dimensionful, and because of that it sets the scale for the masses of
all the known elementary particles [8]. All the elementary particles of
the Standard Model which have mass --- quarks, leptons, $W^{\pm}$, and $Z^0$ 
--- derive these masses from coupling to 
the expectation value of the Higgs. This expectation value, which is
denoted $v/\sqrt{2}$, is determined
by the minimization of the effective potential of the Higgs field, 
$V(\phi) = \lambda
(\phi^{\dagger} \phi )^2 + \mu^2 \phi^{\dagger} \phi$.
Since $\mu^2 < 0$ one has that $v = \sqrt{ \left| \mu^2 
\right|/ \lambda}$. The observed value of $v$ is $246$ GeV. We will
call this value $v_0$, and henceforth the subscript `0' will denote
the value a parameter takes in our universe.
The masses of the weak-interaction gauge bosons,
$W^{\pm}$ and $Z^0$ are given by $v$ times the gauge coupling constants,
which are of order $\frac{1}{2}$. The masses of the quarks and leptons
are given by $v$ times Yukawa couplings which range from about $2 
\times 10^{-6}$ for the electron to about $1$ for the top quark.
The mass of the Higgs particle is $\sqrt{2 \lambda} v$, where $\lambda$ is
as yet unknown, but is roughly of order unity [9]. 
  
The second way in which $\mu^2$ stands out among the parameters
of the Standard Model is its extreme smallness. As noted above,
it is of order $(10 \; {\rm GeV})^2$, or about $10^{-34}$ in natural 
Planckian units. This is to be
compared to the next smallest parameters, the electron Yukawa
coupling, $\lambda_e \cong 2 \times 10^{-6}$, and the QCD
vacuum angle, $\overline{\theta} < 10^{-9}$.
In the simplest grand unified models $\mu^2$ receives contributions 
of order $M_{GUT}^2 \sim 10^{32}$ GeV$^2$ from each of several terms, 
which must therefore cancel to fantastic accuracy. In such 
``fine-tuned" models, a very small change in the other parameters
would disturb this cancellation and cause $\mu^2$ to vary over an
enormous range. (This may be another justification for our
approach here of only considering variations of $\mu^2$ and keeping
the other parameters essentially fixed.) 

Finally, $\mu^2$ stands in contrast to the other parameters of
the Standard Model in that fairly plausible explanations in terms
of symmetry principles or other conventional particle physics
considerations are available to account for their magnitudes.
For example, it is generally regarded as likely that
the relative values of the gauge couplings are the
result of unification of the gauge groups at or below
the Planck scale [10].  Likewise, there are many ideas
for explaining the ratios of quark and lepton masses
in terms of grand unification, family symmetries, horizontal
interactions, radiative hierarchies, or a combination of these [11].
The smallness of the QCD vacuum angle can be explained by
the axion mechanism [12] or by approximate CP invariance [13].
By comparison, the smallness of $\mu^2$ is very hard to account
for in conventional ways. There have been two main approaches to doing this,
``technicolor" [14] and supersymmetry [15]. The technicolor approach is
fraught with difficulties and has fallen into disfavor. The
supersymmetry approach is more promising, but a completely
satisfactory and simple explanation of the smallness of
$\mu^2$ does not yet exist.

It will be assumed, then, that $\mu^2$ can be of either sign
and can vary between $+ M_P^2$ and $- M_P^2$.  
To understand the behavior of $v$ as $\mu^2$ is varied over
this range, one must look at the potential for $\phi$
including the effect of its coupling to the quark-antiquark
condensates (which can ordinarily be neglected).

\begin{equation}
V(\phi) = \lambda \left( \phi^{\dagger} \phi \right)^2 
+ \mu^2  \phi^{\dagger} \phi +
\left( \sum_i \lambda_i \langle \overline{q}_i q_i \rangle
\phi + H.c. \right).
\end{equation}

\noindent
The $\overline{q}q$ 
condensates for light quarks have a value of order $f_{\pi}^3$,
where $f_{\pi}$ is the strong-interaction chiral-symmetry-breaking
scale [8]. ($f_{\pi} \simeq 100$ MeV.) 
For $\mu^2$ negative and
much larger in absolute value than $f_{\pi}^2$, as in our
universe, one can ignore the last term in Eq.(1) and obtain
$v \cong \sqrt{\left| \mu^2 \right|/\lambda} \sim \left| \mu
\right|$. As $\mu^2$ becomes smaller in absolute value than
$f_{\pi}^2$, and of either sign, one can neglect the $\mu^2$
term and obtain $v \sim ( \lambda_t/\lambda)^{\frac{1}{3}} 
f_{\pi} \sim f_{\pi}$. Finally, when $\mu^2$ is positive and
larger than $f_{\pi}^2$ one can neglect the quartic term and
obtain $v \sim  \lambda_t (f_{\pi}^3/\mu^2) \sim (f_{\pi}^3/\mu^2)$.
Note that in the $\mu^2 > 0$ 
world, the longitudinal components of the Weak interaction
gauge bosons come from the ``pions" 
not from the Higgs field, and $M_W \sim g f_{\pi}$.

\section{The $\mu^2 < 0$ universes}

In universes with $\mu^2 < 0$ and greater in absolute value than 
in our universe, one has $v > v_0$. One is dealing, then, with ``large 
$v$" universes. We will examine the possibilities for life in this 
case, our basic assumption being that for life to exist ``complex 
chemistry" must be possible. In our universe, protons and neutrons 
combine to form a variety of nuclei. These are dressed with electrons 
to form atoms which may be bound into a variety of simple and complex 
molecules. It seems plausible, therefore, that in order for life to 
develop it is necessary that a variety of nuclei be (a) stable, 
and (b) formed in either primordial or stellar nucleosynthesis.

One of the things that can go wrong when $v$ becomes larger than $v_0$ 
is that nuclei, or even the protons and neutrons, can become unstable. 
For example, in our universe neutrons can be stable within nuclei; but,
for sufficiently large $v$, as we shall see, the reaction energy for 
neutron decay, $Q = m_n - m_p - m_e$, becomes larger than the binding
energy per nucleon in nuclei of about $8$ MeV. At that point neutrons 
even in nuclei will decay and the only stable nuclei will be protons. 
We argue that such a ``proton universe" would be sterile.

To analyze the stability of nucleons and nuclei in large $v$ universes, 
we must understand how masses and binding energies depend on $v$. We 
turn first to this question.

The quark and lepton masses simply scale with $v$ (ignoring the relatively 
small effect of the logarithmic running of the Yukawa couplings). Thus, 
we take $m_e = 0.5 (v/v_0)$ MeV, $m_u = 4 (v/v_0)$ MeV, and $m_d = 7 (v/v_0)$
MeV. 

We model baryon masses by $m_B = m_q + m_c + m_{em}$, where $m_q$ is the 
sum of the quark masses in the baryon, $m_c$ is the color energy, and 
$m_{em}$ is the electromagnetic energy. Since for the neutron and proton 
the color energy is the same, the neutron-proton mass splitting is given 
by $m_n - m_p = m_d - m_u + m_{em,n} - m_{em,p}$. As long as the
quark masses are small compared to the QCD scale (which will be true for 
$v/v_0$ less than a few hundred) the size of nucleons, and therefore the 
electromagnetic energy, will be relatively insensitive to $v$.
[The size of a nucleon will scale as $\Lambda_{QCD}^{-1}$. For the
dependence of this on $v$ see below.] Thus we
can take $m_{em,n} - m_{em,p}$ to have the same value which it has in our 
universe, namely about $- 1.7$ MeV. Thus we have that $m_n - m_p = 
(3 (v/v_0) - 1.7)$ MeV, and the $Q$ value for neutron beta decay, $Q
\equiv m_n - m_p - m_e$ is $(2.5 (v/v_0) - 1.7)$ MeV.

For $v = v_0$ most of the mass of the $p$, $n$, and $\Delta$ baryons is 
due to color energy. The splitting between the $I= 1/2$ baryons ($n$ and 
$p$) and the $I = 3/2$ baryons ($\Delta$) is in our universe about
$300$ MeV. (Since the lightest baryons will be made purely of $u$ and $d$
quarks, we need be concerned only with isospin and not with flavor $SU(3)$.) 
We will assume that both $m_{1/2}$ and $m_{3/2}$ are proportional to the 
QCD scale, $\Lambda_{QCD}$. $\Lambda_{QCD}$ depends only indirectly
and fairly weakly upon $v$. (This dependence arises because the 
renormalization group running of the strong coupling ``constant", $\alpha_3$, 
depends on quark thresholds, which in turn depend on the quark masses.)
We find that $\Lambda_{QCD} \sim v^{\zeta}$, where $0.25 < \zeta < 0.3$ 
for $-2 < \log (v/v_0) < 4$. Thus we take $m_{3/2} - m_{1/2} \approx 300 
(v/v_0)^{0.3}$ MeV. 
 
Of course, for very large $v$ (larger than a few hundred $v_0$) the quark 
masses will become larger than the color energy, and the proton, neutron, and
$\Delta$ will become non-relativistic bound states in which the color 
energy will go as $\alpha_3^2 m_q$ and thus be proportional to $v$.

The dependence of the nuclear force on $v$ is a much more complicated matter. 
The long-range part of the nucleon-nucleon potential is due primarily
to one-pion exchange, and therefore has a range of $m_{\pi}^{-1}$. As long 
as the $u$ and $d$ masses are small compared to the QCD scale, the mass of the
pion is well approximated by $m_{\pi} \propto ((m_u + m_d) f_{\pi})^{1/2}$. 
Assuming that $f_{\pi} \propto \Lambda_{QCD}$, one has that $m_{\pi}
\sim v^{(1 + \zeta)/2}$. We will take $m_{\pi}$ to go as $v^{1/2}$, which is 
adequate for our discussions. The shorter-range part of the nucleon-nucleon
potential comes from multi-pion exchange, the exchange of heavier states, 
and more complicated effects. It is difficult to estimate how these will 
depend on $v$. However, our qualitative conclusions do not depend
upon this issue.
 
With these general considerations we map out the nature of baryons and 
nuclei in universes where $v > v_0$. 

\begin{itemize}

\item{$v/v_0 = 1$}. In our world, the splitting between isospin multiplets 
is large,  $m_{3/2} - m_{1/2} \approx 300$~MeV~$>> m_q\,,m_{em}$. The 
lightest baryons are thus the proton and neutron. Of these the proton is
lighter because the quark mass splitting ($(m_n - m_p)_{quark mass} = 3$ 
MeV) wins out in competition with the electromagnetic energy splitting
($(m_n - m_p)_{em} = -1.7$ MeV). 

\item{$ 5 \simgt v/v_0 > 1 $}. As $v$ increases the neutron becomes more 
unstable because $m_n - m_p$ increases, and the nuclear potential between 
nucleons gets weaker (since $m_{\pi}$ is getting larger). The combined 
effect is to render nuclei less stable. We estimate that for $v/v_0 \simgt 5$ 
there will be no stable 
nuclei, as the mass excess of the neutron is greater than the nuclear 
binding energy. For $v/v_0 \simlt 5$ a variety of nuclei will 
continue to exist, with fewer and fewer stable isotopes surviving as 
$v/v_0$ increases. 

Even if nuclei are stable there is the question of whether or not they 
may form through nucleosynthesis. Relevant to this question is the
fact that one of the first nuclei to become unstable as $v/v_0$ increases
above 1 will be the deuteron, which even in our universe is 
very weakly bound. This is a particularly important case as all 
primordial and stellar nucleosynthesis ultimately begins with deuterium. 

The critical reaction for decay of the deuteron is $d \rightarrow p + p + 
e^- +\anti{\nu}$ which occurs whenever $B_d < m_n - m_p - m_e
\cong [2.5 (v/v_0) - 1.7] {\rm MeV}$. For the binding energy of the deuteron 
as a function of $v$ we consider two models, both based on the 
knowledge that the deuteron is a weakly bound and rather extended nuclear 
state, sensitive to the long-range pion-exchange component of the nuclear 
potential.

In the simpler model we treat the nuclear potential as a square well with 
a hard core. The hard core mocks up the short range repulsion, whereas 
the square well of depth 35 MeV and width 2 fm represents the one pion 
exchange potential. To model the effects of changing $v$ we decrease the 
width as $(v/v_0)^{-1/2}$ to account for the increase in $m_\pi$. The 
deuteron binding energy can be solved for analytically (see Appendix), 
yielding an approximate relation
\begin{equation}
B_d \cong \left[ 2.2  - a \left( \frac{v-v_0}{v_0} \right) \right]
{\rm MeV},
\end{equation}
where $a \approx 5.5$, for small $v- v_0$. 

The shortcoming of this model is that all the deuteron binding is 
attributed to one-pion-exchange, which probably overstates its 
importance. We therefore tried a more sophisticated approximation
[16], using 
a one-boson-exchange-potential (OBEP) based on deuteron binding and 
scattering phase shifts. This model includes 6 bosons, and also includes 
$s$ and $d$ wave mixing for the deuteron. We varied $m_\pi$ 
proportionally to $v^{1/2}$ (we neglect the scaling of $\Lambda_{QCD}$), 
but it is not clear how the other meson parameters should vary. The 
problem is not well defined, as several of the ``mesons'' do not 
correspond to physical particles but are mockups of the short range 
exchange of two or more pions in channels with the same quantum numbers. 
The masses for these mesons reflect the momentum distribution of the 
multiple pions as much as the mass of the pion itself. Further, the mass 
and coupling parameters are arrived at only after fitting, thus to change 
their relative values in an ad-hoc way can destroy some sensitive 
cancellation, and is of questionable value. Faced with this problem we 
chose to vary the pion mass, but kept all other parameters of the OBEP 
unchanged. Solving for the deuteron bound state we find a nearly linear 
relation between $B_d$ and $v$ of the same form as Eq. 2, with $a 
\approx 1.3$ MeV [16].

For both our models, the deuteron binding energy is explicitly a function 
of $m_\pi$ and implicitly a function of $v/v_0$. The deuteron becomes 
unstable to weak decay for 
\begin{equation}
v/v_0 \approx \frac{3.9 + a}{2.5 + a}
\end{equation}
with $a$ in MeV. For the square-well model and the OBEP model, the 
deuteron is unstable at $v/v_0 = 1.2$~and~1.4 respectively. 

If the deuteron lifetime against weak decay is long enough, then a chain of 
nuclear reactions involving intermediate unstable deuterons may be 
possible. In this case the critical reaction is the strong decay $d 
\rightarrow p + n$, which becomes possible if $B_d < 0$. The corresponding 
values of $v/v_0$ are 1.4 and 2.7 for the square-well and OBEP models, 
respectively.

The anthropic argument in the case where deuterons are unstable is not 
airtight --- there exists the possibility that nuclei may form in 
neutron-rich regions following stellar collapse. Such a scenario would require 
significant rates for three-body processes, or a long-lived deuteron as 
may exist for the regime of $1.4 < v/v_0 < 2.7$ for the OBEP model.

\item{$ 10^3 > v/v_0 > 5$.} For this range of $v/v_0$ nuclei are unstable 
to decay of constituent neutrons, $(A,Z) \rightarrow (A-1,Z) + p + e^- + 
\anti{\nu}$. Hypothetically, there is the possibility of stable proton-rich 
($Z \gg N$) nuclei, but this seems unlikely. In our world the depth of the 
nuclear potential is of order 50 MeV, but the binding energy per nucleon 
in nuclei is only about 10 MeV. The difference is primarily due to kinetic 
energy in the form of nucleon degeneracy energy and coulomb energy due to 
the protons. In a nucleus with $Z \gg N$, with the same value of $A$
and the same nuclear density, the fermi energy will be greater by a factor 
of about $2^{2/3} \cong 1.6$, and the coulomb energy will be greater by 
roughly a factor of 4. Given, also, an expected decrease in nuclear binding 
due to an increase in $m_\pi$ and stable proton-rich nuclei seem unlikely 
even in the absence of inverse $\beta$ decay. 

Even if stable proton-rich nuclei do exist, it will be 
difficult to form them. The most stable nuclei will occur for 
intermediate $A$, but there will be significant gaps in the sequence of 
stable nuclei necessary for nucleosynthesis. For example, either from 
direct experiment or by comparison to other unstable isotopes we may 
conclude that He$^2$, Li$^3$, and Be$^4$, are all strongly unstable in 
our universe, and therefore also in universes where $v>v_0$. It is unclear,
then, how compound nuclei would form if $v/v_0 \simgt 5$. 

Thus, the only stable nucleus will be the proton, and the only element
Hydrogen. We expect such ``proton universes" to be sterile. It is 
interesting that the existence of neutrons close enough in mass to 
the proton to be stable in nuclei plays an important role in making life 
in our universe possible.

\item{$v/v_0 \simgt 10^3$}. For large enough $v/v_0$, the mass difference 
between $u$ and $d$ quarks is greater than the penalty in color energy,
$m_{3/2} - m_{1/2}$,  
that must be paid to have three identical quarks in a baryon. Exactly where 
this occurs is not too important, but at some point $m_p = m_{\Delta^{++}}$. 
Comparing $m_d - m_u \approx 3 (v/v_0)$ MeV to $m_{3/2} - m_{1/2} \approx 
300 (v/v_0)^{0.3}$ MeV suggests that equality takes place at around $v/v_0 
\approx 500-1000$. 

On either side of the critical value there is a range of order 20\% in 
$v/v_0$ where both $p$ and $\Delta^{++}$ are stablized by the electron 
mass. This leads to what we shall call ``proton plus $\Delta^{++}$
universes".

As $v$ is increased above the narrow range of values where $p$ and
$\Delta^{++}$ can coexist, the proton 
becomes unstable to the decay $p \rightarrow \Delta^{++} + e^- + 
\anti{\nu}$, and at this point the only stable baryon is the 
$\Delta^{++}$. We refer to such a universe as a ``$\Delta^{++}$ universe".

It is fairly clear that the $\Delta^{++}$ universes are sterile, as in 
them it seems quite unlikely that two $\Delta^{++}$'s would bind. Since 
$v/v_0$ is large enough that $m_{u,d} > f_{\pi}$, the pion is not a 
goldstone-boson-like particle, and the nuclear force has only a short 
range part. Here, in addition, there is a substantial Coulomb repulsion 
between the $\Delta^{++}$'s. 

If the $\Delta^{++}$'s cannot fuse to form heavier nuclei, then there is 
only a single kind of element, which will have a single $\Delta^{++}$ as 
its nucleus. But this element is chemically equivalent to Helium, and is
therefore chemically inert as well. 
Therefore, in the $\Delta^{++}$ universes one expects neither nuclear nor 
chemical reactions to occur. It is hard to conceive, then, what kind of 
reactions could form the basis of life. 
The ``proton plus $\Delta^{++}$ universes" are only marginally
more interesting. From what we know of their molecular states, it
seems plausible that Hydrogen and Helium alone could not form the
basis of biochemistry.
We conclude, therefore, that the whole range 
from $v/v_0 \simgt 10^3$ to $v/v_0 \sim M_P/v_0 \sim 10^{17}$ can be 
excluded anthropically.

\item{$1 > v/v_0 $} Finally, although it is not strictly an issue for the 
hierarchy problem, we examine the nuclear consequences of $v<v_0$ in 
universes where $\mu^2<0$. As $v$ decreases from $v_0$, the neutron 
becomes stable, then $m_n = m_p$, followed eventually by a region where 
the proton is unstable to decay $p\rightarrow n+e^++\nu_e$. The stability 
criteria are determined using a $v$ dependent electron mass $m_e = 
\lambda_e v$.

One result of increasing neutron stability is to increase the primordial 
nucleosynthesis yield of He$^4$. Once the neutron is lighter than the 
proton we can no longer reliably estimate the results of nucleosynthesis.
With no coulomb barrier to suppress reactions, we anticipate that all 
single nucleons are bound into compound nuclei, but we cannot calculate 
the distribution of heavy elements. The value of $m_p - m_n$ never gets 
larger than 1.7 MeV, so it is never large enough to destablize nuclei 
in a manner similar to neutron decay in the $v/v_0 > 5$ regime. These 
compound nuclei are therefore stable and we see no reason why values 
of $v/v_0 < 1$ could not support life.

\end{itemize}

\vspace{0.5cm}

\section{The $\mu^2 > 0$ universes}

For $\mu^2 < 0$ the anthropic argument based on viable chemistry worked 
very effectively. The situation is not as simple for $\mu^2 > 0$.
In this case, $v \sim f_{\pi}^3/\mu^2$, and therefore for 
$\mu^2 \simgt \left| \mu_0^2 \right| \sim (10^2 {\rm GeV})^2$
one has $v \simlt 10^{-9} v_0$, and all the quark and lepton masses
are extremely small. For example, $m_e \sim 5 \times 10^{-4} 
{\rm eV} (\mu_0^2/\mu^2)$. Also, symmetry breaking of $SU(2) \times U(1)$ 
is driven by quark condensates, so that $M_W \sim f_\pi$. These facts 
have dramatic consequences for the chemical energy scale of life, the 
structure of elements and stellar evolution; all of which play a 
role in the genesis of life. 

Our critical assumption is that life
requires the chemistry of complex molecules. A typical
biochemical energy is $E_{chem} = \epsilon
\alpha^2 m_e$, where $\epsilon$ is a numerical factor, which
in our universe is of order $10^{-3}$.
(This gives, not coincidentally, $E_{chem} = 300$K, the average
surface temperature of the earth.) What $\epsilon$ would be in 
the bizarre small-$v$ universes we shall consider is hard to say.
We shall keep it as a parameter in our formulae. We think it
unlikely that it is large compared to one. 
The crucial point is that $E_{chem}$ is proportional to $m_e$,
which in $\mu^2 > 0$ universes is tiny compared to its value in our 
universe. 

It is clear that chemical life cannot emerge until at least
the time, which we will call $t_{chem}$, when the temperature 
of the cosmic background radiation cools below $E_{chem}$.
To put it picturesquely, there is the problem of life being fried
by the cosmic background radiation. When this radiation has
temperature $T \sim E_{chem}$, the matter density will
be $\rho (t_{chem}) \sim \eta_B m_p E_{chem}^3$, where $\eta_B$ is
the baryon to photon ratio. We assume $\eta_B$ to be set
by physics that is insensitive to the value of $\mu^2$, such
as the interactions of a grand unified theory, and therefore to have
a value similar to that in our universe, namely
$10^{-10}$. Using $t_{chem} \sim M_P/
\rho( t_{chem})^{\frac{1}{2}}$, $E_{chem} = \epsilon \alpha^2 m_e$,
and $m_e \sim m_{e0} (f_{\pi}^3/\mu^2 v_0)$, one derives

\begin{equation}
t_{chem} \sim m_p^{-1} \left[ \left( \frac{M_P}{m_p} \right)
\left( \frac{m_p}{m_{e0}} \right)^{\frac{3}{2}}
\left( \frac{v_0 \left| \mu_0^2 \right|}
{f_{\pi}^3} \right)^{\frac{3}{2}} \alpha^{-3}
\eta_B^{-\frac{1}{2}} \right] \epsilon^{-\frac{3}{2}} 
\left( \frac{\mu^2}{\mu_0^2} \right)^{\frac{3}{2}},
\end{equation}

\noindent
or

\begin{equation}
t_{chem} \sim  10^{19} {\rm yrs}
\; \epsilon^{-\frac{3}{2}} 
\left( \frac{\mu^2}{\mu_0^2} \right)^{\frac{3}{2}}.
\end{equation}

\noindent
By this time one of several disasters may occur which
prevent the emergence of life. We
will discuss two of these: the burning out of stars, and the
decaying away of all baryons. Before we do so, however, we must
reexamine the assumptions that went into Eq. (5). First, we note 
that this is a lower limit for $t_{chem}$ as other contributions to 
the radiation density (e.g. baryon decay, see below) may be greater than 
the primordial cosmic background radiation. Second, we have 
assumed that chemistry is based on electrons, an assumption which we now 
examine.

It is not hard to show that
if we replace $m_e$ by either $m_{\mu}$ or $m_{\tau}$ in the formula
for $t_{chem}$ the resulting time is still short compared to
the lifetimes of the $\mu$ or $\tau$. This is because the $\beta$-decay
lifetime of a lepton goes as one over the fifth power of its mass, and
therefore as $v^{-5} \sim (f_{\pi}^3/v_0 \mu^2)^{-5}$. (Recall that
the Fermi constant, $G_F \sim f_{\pi}^{-2}$ in this limit rather than
as $v^{-2}$.)   

Since $\mu$'s and $\tau$'s are stable on the relevant time scales,
it is possible that the valence leptons of biochemistry would be
some lepton other than (or in addition to) electrons. This cannot be
ruled out, and in fact whether it is so depends on the details
of baryo/leptogenesis. Whatever the mechanism of baryo/leptogenesis,
sphaleron processes will certainly have the effect of reshuffling
baryon and lepton numbers in such a way that there will be non-zero
values of all three lepton numbers. Further, there is no reason for all 
three lepton numbers to be negative, even though sphaleron processes will 
try to minimize $B+L$. Let us consider two cases to
see some of what is possible. (1) If primordially only $\tau$-number was
produced, and it was negative, then sphaleron-processes, weak scattering
processes, and beta decay would lead to a distribution with a net 
positive charge in $\tau$s, $Q_\tau > 0$. Additionally, baryon's (see 
below) tend to be positively charged, so $Q_B>0$. These charges are 
balanced by $Q_e<0$ and $Q_\mu<0$. 
This means that baryons and $\tau^+$'s would become clothed with 
$e^-$'s and $\mu^-$'s. Since the $\mu^-$'s would be much more tightly
bound, chemistry in this case would be electron chemistry.
(2) If, on the other hand, primordially only electron number were 
produced, and it were negative, then the baryons would end up being clothed
exclusively with $\tau^-$'s --- as long as $N_\tau > Q_B$. In any baryonic 
atom, a $\tau$ could replace a $\mu$, with a gain in binding energy. The 
remaining $\tau$s and $\mu$s would bind, albeit more weakly, with 
electrons. 

If the chemistry relevant for life is dominated by lepton $l^-$, then one 
simply multiplies the formula for $t_{chem}$ by 
$(m_{e0}/m_{l0})^{\frac{3}{2}}$. This result is to be compared
with the baryon lifetime. If $\tau_B > t_{chem}$, we then 
consider the structure of nuclei, and stellar evolution.

\vspace{0.5cm}

\noindent
{\bf (i) The decay of baryons.}

The unification of gauge couplings suggests grand
unification at a scale of order $10^{16}$ GeV, and therefore
the existence of gauge bosons of that mass whose exchange leads
to violation of baryon number.
But even if there is no unification at that scale, it is 
plausible to suppose
that at least at the Planck scale there will be states whose
exchange violates baryon number. Thus we will parametrize
the baryon decay rate as $\Gamma_B = m_p^5/M^4$, where $M$
is assumed to lie between $10^{16}$GeV and $10^{19}$GeV.

As mentioned above, if baryons are not stable the radiation density may 
be dominated by the energy released in their decay. We equate the age of 
the Universe, $t \sim \simeq M_P \rho^{-\frac{1}{2}}$, to the baryon 
lifetime. At this time the radiation density is given by the relation 
$\rho_{rad} = f \rho_B$, where $f$ is a numerical factor that we take to 
be about $0.3$ [17]. This gives $T_{rad} \simeq (f/g_{rad})^{\frac{1}{4}} 
(\Gamma_B M_P)^{\frac{1}{2}} \simeq 0.3 (\Gamma_B M_P)^{\frac{1}{2}}$. 
Requiring that this be less than $E_{chem} = \epsilon \alpha^2 m_e \sim 
\epsilon \alpha^2 m_{e0} (f_{\pi}^3/\mu^2 v_0)$, and using the 
parametrization of $\Gamma_B$, yields the constraint

\begin{equation}
\mu^2 \stackrel{_<}{_\sim} \epsilon \alpha^2 \left(
\frac{m_{e0} f_{\pi}^2 M_P^{\frac{3}{2}}}{v_0 m_p^{\frac{5}{2}}}
\right) \left( \frac{M}{M_P} \right)^2 \cong \epsilon (4 \times
10^5 {\rm GeV})^2 \left( \frac{M}{10^{16} {\rm GeV}} \right)^2.
\end{equation}

\noindent
Thus if $M \sim 10^{16}$GeV, the constraint that baryons
still exist when the universe is cooled down to $E_{chem}$
tells us that $\mu^2$ must be at least $27$ orders of magnitude
smaller than the natural scale of $M_P^2$. The 
value in our universe is about $33 \pm 1$ orders of magnitude 
smaller than the natural scale. If biochemistry is controlled
by $\mu$'s or $\tau$'s, then the above constraint on $\mu^2$
would be weakened by a factor of $m_{\mu}/m_e \approx 200$,
or $m_{\tau}/m_e \approx 3,000$.  

There are a number of considerations that may actually strengthen
the bound on $\mu^2$. In the first place, we have neglected 
the possibility that baryons might decay
predominantly into channels that are blocked in our universe
by the heaviness of the higher generation quarks and
leptons. For example, in some supersymmetric theories the
exchange of superheavy colored Higgsinos would produce
very fast proton decay (in our universe) if it were not
for the fact that the final states include flavors 
heavier than the proton.

Even more interesting is the possibility that baryons could decay
{\it via} sphalerons [18]. Since in $\mu^2>0$ universes the weak 
interactions are broken not at $v_0$ but at $f_{\pi}$, 
non-perturbative weak baryon decay would not be as strongly
suppressed as in our universe.

Finally, there is the question of baryogenesis. If it occurs
through standard ``drift and decay" mechanisms [19] involving
grand unified interactions, then one would expect the asymmetry
to be essentially independent of $\mu^2$. But if baryogenesis
takes place at the Weak scale in our universe [20], it is probable that
in the $\mu^2> 0$ universes the asymmetry would be 
very different. 

\vspace{0.5cm}

\noindent
{\bf (ii) The structure of the elements}

With the Higgs system not generating a vacuum expectation value, the 
breaking of the Standard Model occurs through the Goldstone bosons of  
the strong interactions.  The $W$ mass is of order 50 MeV. However, the 
quark masses are very small since they are sensitive only to the Higgs 
vacuum expectation value. With six essentially massless quarks, the 
lightest baryon multiplet will be a 70-plet, with 27 members which are 
neutral, 27 which have charge +1 and eight each with charge -1 and charge 
+2. The lightest meson multiplet  starts out with 35 members, but three 
are ``eaten" to become the longitudinal gauge bosons, leaving 32 
pseudoscalar mesons, which are the Goldstone bosons of the dynamically 
broken chiral SU(6) symmetry.

Electroweak interactions will split the multiplets and our estimates
for these effects are based on the understanding of electromagnetic
mass splittings of the observed hadrons. For the Goldstone bosons the
mass splittings can be understood using effective Lagrangians with
chiral symmetry, which in the world under consideration would involve a 
$SU(6)_L \times SU(6)_R$ chiral invariance. 
The purely lefthanded interactions of the charged $W$'s will not
produce masses for the Goldstone bosons, since such interactions
lead to an
effective chiral lagrangian which involves derivatives [21], and hence
vanishes at $p^2 = 0$. However vectorial interactions do generate
masses, in analogy to the electromagnetic mass shifts of pions and
kaons. The
electromagnetic and $Z^0$ interactions will have such vectorial
effects, and display 
an $SU(3)_u  \times SU(3)_d$
invariance for separate rotations of up-type and down-type quarks.
By a generalization of Dashen's theorem [22], these interactions 
will leave the
16 neutral Goldstone bosons 
massless, while giving a common mass to all 16
charged mesons. Using the observed pion mass splitting we can estimate
that the charged mesons will have a mass of about $35$ MeV. 
The neutral mesons will develop very small masses due to chiral symmetry 
breaking, $m_\pi \propto (m_q f_\pi)^{1/2} \approx \lambda_q^{1/2} 
f_\pi^2/\mu$. Even for $\mu$ as small as 100 GeV, some meson masses will be 
less than a KeV.
The tiny mass of the neutral mesons implies that nucleons will
have a long-ranged neutral mesonic cloud, and that the nuclear force
will have a correspondingly long-range potential.

The baryon mass splittings are less amenable to an analysis based only
on symmetry, and we must include quark model ideas about the effects
of electroweak interactions. The short-distance electroweak effects
preserve the quark chirality, which implies that they will not
generate quark mass shifts. The longest range effects of the vectorial
interactions will be
proportional to the square of the baryon's electric or weak charges. 
Describing these by the electric charge Q and the strong isospin t, and
ignoring the masses of the $W,Z$, we would expect that the lightest
baryons should be the 19 $(Q,t) = (0, {1 \over 2})$ states, a group
which contains the neutron. With splittings of order an MeV, the
19 $(Q,t)=(1,{1\over 2})$ charged baryons (containing the proton and
related states) could likely be the next grouping. Hyperfine
interactions and quark masses
could further split the states within these groups.
All of 
the baryon and meson states which are shifted up in mass 
are unstable and can decay down to the ground states via 
weak or electromagnetic interactions, although perhaps with very long 
lifetimes.

Given these building blocks we outline the nature of 
nuclei. In our world, consisting of just protons and neutrons, any 
substantial nucleus has comparable numbers of neutrons and protons, as 
dictated by a desire to minimize the fermi energy. There is a tendency to 
have fewer protons so as to simultaneously lower the coulomb energy. In 
the $\mu^2>0$ worlds, there are many species of neutral and charged 
nucleons, so the effects of degeneracy energy will only come into
play for larger 
nuclei. At the same time, for modest size nuclei ($R$ of a few fermi) the 
weak bosons are effectively long range, $m_W R \approx m_Z R < 1$, so the 
one must take into account weak-interaction energy as well as
electromagnetic coulomb energy.

A nucleus consisting of $N$ baryons of type $(Q, t_3) = (0, -\frac{1}{2})$
and $Z$ of type $(Q, t_3) = (1, \frac{1}{2})$ will have strong isospin 
$t_3 = \frac{1}{2} (Z - N)$ and weak hypercharge
$Y/2 = \frac{1}{4} N + \frac{3}{4} Z$. (Since right-handed quarks have
vanishing weak isospin, the effective weak isospin of a nucleon is half
of its strong isospin, and $Y/2 = Q - t_3/2$.) The coulomb energy
will have contributions proportional to $(g_1^2 (Y/2)^2 + g_2^2 t(t + 1)/4)$.
For large $t$ we can approximate $t(t+1)$ by $t^2$. Clearly, this is 
minimized by making $t$ as small as possible, namely
$t = \left| t_3 \right|$. This allows us to express the coulomb energy 
in terms of $N$ and $Z$, and minimizing with respect to $Z$ subject
to the constraint that $Z + N = A$, one finds easily that
$$
Z=\frac{A}{2}\frac{g_2^2 - g_1^2}{g_2^2 + g_1^2} = 
\frac{A}{2}(1-2\sin^2(\theta)) \approx \frac{A}{4}.
$$
This relation holds for intermediate $A$; for large $A$ the range of the 
weak gauge bosons will saturate and eventually we must include 
degeneracy effects (filling fermi levels) as well as other nucleon states, 
while for small $A$ we must include the 
nucleon mass differences.

Coulomb energy may cause nuclei larger than some critical size to be
unstable to spontaneous fission, as in our 
universe. However, above the value of $A$ for which the weak force saturates 
(which we estimate to be on the order of a few hundred) 
the ratio $Z/A$ is determined predominantly by minimizing
electric coulomb energy rather than weak-interaction coulomb energy.
This would lead (in these $\mu^2 > 0$ universes) to large nuclei
having $Z \ll A$.
The smallness of $Z/A$ may well mean that there is no maximum size
set to nuclei by spontaneous fission. The situation is 
complicated by the longer range of the strong nuclear force 
and the effects of degeneracy energy 
with many degrees of freedom. Therefore, we cannot say with certainty 
whether or not there is a maximum nuclear size in the $\mu^2 > 0$ worlds.

Similarly, the spectrum of nuclei that result from primordial 
nucleosynthesis is not certain. Early stages of nucleosynthesis will 
occur with neutral baryons combining to form light nuclei, but as nuclei 
grow in size and charge, it is not clear to us whether or not coulomb barriers 
will result in a termination of nucleosynthesis at small or modest 
nuclei, or whether nuclear burning will ``run away" to give only
very massive nuclei. The difficulty lies in estimating how screening 
and thermal contributions to meson masses will affect the long-range 
nuclear potential in its competition
with the coulomb barrier between light nuclei.

With the uncertainty in these issues, it seems possible that light nuclei 
with small charge may exist and provide a basis for chemical life. It is 
also possible that nucleosythesis will result in a small number of low-charge, 
superheavy nuclei, which does not seem conducive to the development of 
chemical life.

\vspace{0.5cm}

\noindent
{\bf (iii) Stars burning out.}

Even if an interesting mix of elements develops during nucleosynthesis, 
the time $t_{chem}$ is so large that it is natural to wonder if there 
would be any stars left by the time biochemistry could take place in the 
small-$v$ universe. On the other hand, with extremely small lepton 
masses, stellar cooling may be so obstructed that stars never contract to 
the nuclear burning phase at all. We give an overview of these issues.

Stars are supported by either gas pressure, radiation pressure, or 
degeneracy pressure. If the support is either gas, radiation, or 
degeneracy of a relativistic species then as the star cools (loses 
energy) it contracts and heats up ($T$ increases). If supported by 
degeneracy of non-relativistic fermions, then it cools but does not 
contract. A system of the first kind contracts until it becomes hot enough 
for nuclear burning to proceed against the coulomb barriers of the 
nuclei. Once the fuel is gone contraction continues until either another 
fuel burning stage is reached, the object is supported by 
non-relativistic degeneracy pressure (white dwarf or neutron star), or 
the object contracts within its Schwarzschild radius and disappears as a 
black hole. Systems supported by degeneracy pressure before nuclear 
burning cool into planets. 

Whether a cloud of gas turns into a star or a planet depends on its 
initial mass and (to a lesser extent) composition. Initially the cloud is 
non-degenerate and is supported by gas pressure. As it  
contracts degeneracy pressure increases as $R^{-5}$ while gas or radiation 
pressure increase as $R^{-4}$. When the star reaches a size $R_d$ 
degeneracy pressure will halt further contraction. From dimensional 
analysis,
$$
R_d \sim \frac{N^{1/3}}{M_* m_f} \,,
$$
where $M_* = (G_N m_B^2)^{3/2} N \approx M/M\solar$, $N$ is the 
number of baryons in the star, and $m_f$ is the mass of the degenerate 
fermion [23].

The temperature at this point is
$
T_d \sim M_*^2 m_f\,,
$
and the fermi momentum of the degenerate fermions is
$
k_{f_d} \sim M_* m_f \,.
$
If $T_d$ is greater than the temperature necessary for nuclear burning, 
$T_N \approx$ 1 KeV [24],
then a star is born before degeneracy occurs. In our 
world, $m_f = m_e$ and, after including numerical factors, $T_d > T_N$ 
for $N_B > .08 N\solar$. Low mass stars develop, burn, and then turn into 
white dwarfs. In the $\mu^2>0$ world, the lepton may be $e$, $\mu$, or 
$\tau$, but in any case $m_f<1$ eV. $T_d$ will therefore be too cool to 
support nuclear burning. Protostars with $M_* < 1$ turn into planets.

If $M_* > 1$ the leptons become relativistic before they become 
degenerate, and the collapse cannot be halted until nuclear fusion takes 
place. After a burning phase contraction continues until either 
another nuclear burning phase occurs or the core contracts inside its 
Schwarzschild radius and a black hole results. If $M_*$ is not too much 
greater than 1, a neutron star may form.

Since $M_* > 1$ for nuclear fusion, stellar burning will take place in a 
plasma of relativistic leptons and anti-leptons. This plasma is not 
degenerate: the degeneracy parameter is $k_f/T \approx 1/M_*$. As the 
leptons are relativistic opacities for photons will be large, dominated 
by the photon-lepton cross-section $\alpha^2/T^2$. On the other hand 
neutrino interactions are also much larger, so neutrinos dominate 
stellar cooling in the nuclear burning phase.

At temperatures less than $M_W$ cross-sections for neutrino pair 
production, scattering from leptons or baryons, absorption, etc, will be 
of order $T^2/f_\pi^4$. For stars of $M_*=1$ the density of baryons and 
thermal leptons will be comparable, but for larger stars thermal 
particles will dominate, so we may estimate mean free paths and 
emissivities from 
thermal pair processes. Mean free paths for weak interactions are 
$\lambda \sim f_\pi^4/T^5 \approx 10^{12}/T_{KeV}^5 $cm ; ie, solar type 
stars are likely to cool by volume emission or have a very deep neutrino 
sphere. The energy loss rate per lepton from pair annihilation to 
neutrinos is $\approx 10^{-8} T_{KeV}^6$ GeV s$^{-1}$; ie, solar mass 
stars cool on time scales of roughly a year, and larger stars in much less
time. This is very much less than $t_{chem}$. 

The next issue is star formation. Of particular concern 
are the cooling rates during collapse, when temperatures are too low for 
efficient neutrino emmission, and cooling is 
regulated by the photon opacity either in the interior or at the surface. 
Opacities are determined by the chemistry of the lightest charged lepton. 
For us this is the electron mass, although in some scenarios the 
active species will in fact be positrons bound in either $\mu$ or 
$\tau$-onium.

There are two stages in the collapse of a cloud to the point of nuclear 
ignition. At first, temperature gradients are not large and convection 
provides the energy transport. The photosphere is essentially held at a 
fixed temperature $T_e \approx \alpha^2 m_e$ at which material is no 
longer ionized and the opacity drops. The luminosity is $L \sim R^2 
T_e^4$. Cooling is initially very fast but slows as the star shrinks. 
Eventually, temperature gradients increase to where radiative transport 
is effective and convection is cut off. At that point the cooling time 
scale is the Kelvin-Helmholtz time for photon diffusion to the surface.
$$
t_{KH} \sim \frac{R^2}{\lambda} \sim \frac{N}{R}\sigma 
\approx \frac{10^{38} \alpha^2 T}{T^2 + m_e^2}\,,
$$
where $\lambda$ is the mean free path and $\sigma$ is the cross-section 
for photon-lepton scattering. To derive the last relation: note that if $T<m_e$ 
then $\sigma \sim \alpha^2/m_e^2$ but for higher temperatures $\sigma \sim 
\alpha^2/T^2$, and that $RT \approx 10^{19} M_*$ is 
roughly constant during collapse. The cooling time is dominated by the 
epoch when $T=m_e$, or
$t_{cool} \approx 10^{34}/m_e \approx 10^{17} \mu^2/\mu_0^2$ yr.
This is less than $t_{chem}$ (see Eq. (5)), 
but not by so much that the 
energy from stellar burning may not be available for life to form. 

Thus, within this crude treatment of stellar evolution, stars are
expected to form slowly, but burn nuclear fuel very quickly. The actual
stage of nuclear burning seems too short to benefit life, but there are 
other possible energy sources than surface heating of planets by stars.
For example, planets may be volume 
heated by radioactivity, residual gravitational energy, baryon decay, or 
even absorption of the background of neutrinos produced by stellar 
burning. The energy flow to the surface would in principle be usable for 
the evolution of life.

\section{Quark Masses}

We have been discussing theories in which the Higgs mass parameter
$\mu^2$ can vary in different regions of the universe, under the assumption that
the Yukawa couplings do not change from the values that they have in our portion
of the universe.  It is possible that
the underlying theory also allows the values of the Yukawa couplings to 
vary in different regions of the universe. In this section,
 we discuss some of the
possible implications of this situation. However, we 
stress that without knowing the details of the fundamental 
theory we do not know whether these masses are in fact subject 
to  variation or whether they occur in fixed ratios due to 
some other mechanism.
For example, we can see no anthropic argument which would 
force neutrinos to be as light as they seem to be (if indeed they have
a mass). 
However, a ``see-saw" mechanism [9] would make them very light 
automatically, and we would not need consider anthropic reasoning.

The masses of the quarks and leptons of the second and third 
generations have very little impact on the particles and reactions 
which occur naturally in our universe. Therefore anthropic arguments 
would not place constraints on their masses. In an ensemble of anthropically 
allowed universes, these masses could be randomly distributed.  In
practice, 
the observed masses appear to be distributed without any obvious 
pattern, and randomness appears as good an ``explanation" as anything else.

The masses of the light quarks are constrained by the physics which
has been  
discussed above. In particular, the requirement that the 
deuteron exist yields an upper bound on the sum of the up and down quark masses,

\begin{equation}
{(m_u + m_d )_{max} \over (m_u  + m_d)_{real}} \le 2
\end{equation}
 
If the $d$ quark mass was lighter, or the both the masses were small, 
the proton would be heavier than the neutron, and could decay into it.
Hydrogen would then not exist as a stable atom (at least for small 
values of the electron mass). However, the complex elements would 
still exist, and there would seem to be sufficient building blocks 
for  some form of life. Thus, we will use the above inequality as 
the sole anthropic constraint on the light quark masses.

The anthropic constraints on the electron mass come from other 
sources. An electron mass larger that the binding energy of a 
nucleon, around 10 MeV, would lead to the decay of atoms, through 
the process $e^- + p \to n + \nu$, with the final neutron ejected 
from the nucleus.  However, stronger constraints can be obtained 
in nucleosynthesis.  If the electron mass is higher than the
temperature 
at the time of nucleosynthesis, the electrons will have
all disappeared, converting via $e^- + p \to n + \nu$ leaving 
only neutrons behind. Likewise, the reactions which would 
burn neutrons, such as $n + n \to d + e^- + \bar{\nu}_e$ 
which would be the neutron equivalent of the start of the 
pp cycle,  use weak interaction transitions and 
would be shut off if the electron mass is large. The precise 
constraint depends on the neutron and proton masses, but is 
generically in the range of a few MeV.

\section{Conclusions}

In a universe with domains which can have 
different values of some of the underlying parameters, 
life may only be able to develop in some of those domains. 
If this is the case, we would expect that the parameters 
of our domain should be typical of anthropically 
allowed range.  If the anthropic principle accommodates 
such a large range that our values of the parameters are 
unnaturally small within this range, then the anthropic 
principle fails to help us understand the sizes of these 
parameters. However, 
we have found that within the overall structure 
of the Standard Model there is a relatively small acceptable 
range for the Higgs parameter $\mu^2$
and the light quark masses. 
The physical values of these parameters are 
quite typical of this range, raising
the possibility that the anthropic principle could 
be an ``explanation" of these magnitudes.

The arguments behind this conclusion are summarized in 
Fig. 1. For $\mu^2$ negative, as in our universe, it seems that
the whole range of values for the vacuum expectation value 
from $M_P$ down to about
$5$ (or perhaps even down to $1.2$)
times the value in our universe can be excluded.
For most of that range (down to about $10^3 v_0$)
the universe would consist of sterile, Helium-like atoms
whose nuclei were $\Delta^{++}$. There would be essentially
no reactions either chemical or nuclear. For the lower
part of the excluded range, there would be virtually no
nuclei other than protons, and the $pp$ and $pn$ processes that
are needed for nucleosynthesis would be endothermic as the deuteron
would not be stable. 

For positive values of $\mu^2$ the condition that 
baryons still exist at the time biochemistry becomes possible
forces $\mu^2$ to be many orders of magnitude smaller than
the ``natural" Planck scale. ({\it Cf.} Eq. (4).) Our arguments for
smaller values of $\mu^2$ are less certain. It may be that long range 
nuclear forces cause all baryons to clump into superheavy nuclei with 
small charge, which doesn't appear to be promising for life. If these 
forces are screened in a mesonic plasma, then
light nuclei will continue to exist, and can burn in stars, and stars may 
ignite at sufficiently late times to fuel life. Individual stars, 
however, will be extremely short lived compared to the cosmological time 
scales. If life is to develop in such a universe, the energy source 
is not likely to be the photoluminescence of an individual star.

One can thus plausibly argue that for life to exist
the $\mu^2$ parameter has to be negative and has to
be close to the value it has in our universe. 

One of the interesting features of this argument is that
it explains --- as no other approaches do at present ---
the curious fact that the Weak scale is near to the 
QCD scale. In order for protons and neutrons (rather
than $\Delta^{++}$) to be the lightest baryons, $m_d - m_u$
has to be less than the
chromodynamic energy which splits the baryon decuplet
from the octet.
For the deuteron to have large enough binding
energy to save neutrons from virtual extinction
in the early universe and also to allow the $pp$ reactions
to be exothermic, the pion has to have a long
Compton wavelength compared to the nucleon, and this in turn means
that the $u$ and $d$ masses have to be not only less
than but small compared to the QCD energy scale by about
the amount seen in nature.
This provides a possible resolution to the ``fine-tuning problem'' ---
in an ensemble of different domains of the universe, the Higgs mass 
parameter will occasionally fall into the anthropically allowed region
without having to be fine-tuned in general. If the cosmological
constant is confirmed to have a non-zero value close to present
estimates, and no new physics is found in the TeV energy region, we
may be faced with de-facto evidence for the presence of fine-tuning.
In such a situation, the anthropic or multiple-universe considerations
become highly attractive.  

Finally,  let us comment on the ability of these 
ideas to be tested. Negatively, we can say that if the
Weak scale is what it is for anthropic reasons, there would be no
need to invoke supersymmetry or technicolor or other structure at
the Weak scale to make the fine-tuning ``natural" [1]. If no such
structure is found, then, it would be a point in favor of anthropic
explanations; indeed, in that case there would be few if any 
alternatives to an anthropic explanation.
Positive evidence is harder to come by. Of course, we are not able to 
explore other domains in the universe. 
However, theories which generate multiple domains may be 
testable by other, more conventional means. For 
example, the community is hoping to be able to 
test the details of inflationary theories through 
cosmological measurements. Likewise, direct physical 
experimentation has the potential to eventually sort out 
the correct underlying theory. Through standard means 
we may be able to learn if the fundamental theory in 
fact produces multiple domains, in which case 
anthropic considerations automatically become 
relevant. Until the time that this happens, our conclusion 
must be modest: 
the observed values of the mass 
parameters are reasonably typical of the
anthropically allowed ranges.

\section*{Appendix}

Let us make a simple qualitative estimate of the deuteron binding energy.
A well-known pedagogical model of the deuteron involves a square well
of depth $V_0 = 35$ MeV and range $R = 2$ fm, with a hard core
of radius $r_0$.
Let us neglect the $D$-wave component, so that the $S$-wave solutions
are

\begin{equation}
u_{<}(r) =  r \psi_{<} (r) = A \sin \kappa ( r- r_0), \; \; r_0 \leq r \leq R,
\end{equation}

\noindent
and

\begin{equation}
u_{>}(r) = r \psi_{>} (r) = B e^{-\gamma r}, \;\; r \geq R,
\end{equation}

\noindent
with $\kappa \equiv \sqrt{m(V_0 - B_d)}$, and $\gamma \equiv 
\sqrt{mB_d}$. The boundary condition is therefore

\begin{equation}
\kappa \cot \kappa (R - r_0) = - \gamma.
\end{equation}

\noindent
There will be a range, $R_c$, at which the binding energy
goes to zero, {\it i.e.}

\begin{equation}
\kappa_c \cot \kappa_c (R_c - r_0) = 0, \;\; \kappa_c \equiv
\sqrt{mV_0}.
\end{equation}

\noindent
If we let $\delta R = R - R_c$, we can solve for the present value
of $R$ {\it via} a Taylor series. To first order in $\sqrt{B}$
we have 

\begin{equation}
\cot \kappa_c ( \delta R + (R_c - r_0)) \simeq - \sqrt{\frac{B}{V_0}} 
\Rightarrow  \tan \kappa_c \delta R \simeq - \kappa_c \delta R
\simeq - \sqrt{\frac{B}{V_0}}.
\end{equation}

\noindent
Now, the outer range of the potential is determined by
the pion mass, $R \propto 1/ m_{\pi}$. If we search for the
value of the pion mass at which the deuteron becomes unbound,
we equate

\begin{equation}
\frac{\delta m_{\pi}}{m_{\pi}} =\frac{\delta R}{R} = 
\sqrt{\frac{B}{V_0}} \frac{1}{\kappa_c R} \approx 0.2.
\end{equation}

\noindent
Thus, according to this calculation, only a 20\% increase
in the pion mass would cause an unbound deuteron.
And since the pion mass and the quark masses very nearly obey
the relation 

\begin{equation}
m_{\pi}^2 = (m_u + m_d) \frac{\langle 0 \left| 
\overline{\psi} \psi \right| 0 \rangle}{F_{\pi}^2}
\propto v.
\end{equation}

\noindent
a 20\% increase in the pion mass corresponds to
a 40\% increase in $v$, or a factor of $2$ increase in
$\left| \mu^2 \right|$.

We shall parametrize  the vacuum expectation value as

\begin{equation}
B_d \cong \left[ 2.2 - a \left( \frac{v-v_0}{v_0} \right)
\right] {\rm MeV},
\end{equation}

\noindent
for small $v- v_0$, where $a$ is some positive constant.
The square-well calculation gives $a \simeq 5.5$.

\section*{References}

\begin{enumerate}
\item G. `t Hooft, in {\it Recent Developments in Gauge Theories},
Proceedings of the Carg\`{e}se Summer Institute, 
Carg\`{e}se, France, 1979, edited by G. 't Hooft {\it et al.}, NATO ASI
Series B: Physics Vol. 59 (Plenum, New York, 1980);
H. Georgi and S.L. Glashow, {\it Phys. Rev.} {\bf D6}, 2977 (1972).
\item For a comprehensive review see S. Weinberg, {\it Rev. Mod. Phys.}
{\bf 61}, 1 (1989).
\item S. Weinberg, {\it Phys. Rev. Lett.} {\bf 59}, 2607 (1987);
S. Weinberg, astro-ph/9610044, to be published in the proceedings of the
conference {\it Critical Dialogues in Cosmology} at Princeton University, 
June 1996; H. Martel, P. Shapiro, and S. Weinberg, astro-ph/9701099.
\item B. Carter, in I.A.U. Symposium, Vol 63, ed by M. Longair (Reidel,
Dordrecht, 1974);
J. Barrow and F. Tipler, {\it The Anthropic Cosmological Principle}
(Clarendon Press, Oxford, 1986);
B. J. Carr and M.J. Rees, Nature {\bf 278}, 605 (1979);
A. Vilenkin, Phys. Rev. Lett. {\bf 74}, 846 (1995).
\item A. Linde, {\it Phys. Lett.} {\bf B129}, 177 (1983);
{\it ibid.} {\bf B175}, 395 (1986), {\it ibid.} {\bf B202},
194 (1988), Phys. Scri.,{\bf T15}, 169 (1987).
\item E. Gildener and S. Weinberg, {\it Phys. Rev.} {\bf D13}, 3333 
(1976); E. Gildener, {\it Phys. Rev.} {\bf D14}, 1667 (1976).
\item Changes in $\mu^2$ can indirectly make small changes in the 
dimensionless parameters of the Standard Model through the Renormalization
Group. These are generally negligible. The cases where they are not
will be indicated explicitly.
\item Even though $\mu^2$ is the only dimensionful parameter that
appears explicitly in the Lagrangian of the theory, a scale, 
$\Lambda_{QCD}$, is
dynamically produced by the strong interactions through ``dimensional
transmutation". This sets the scale for chiral-symmetry breaking and
for the baryon masses. However, the masses of the elementary particles
(quarks, leptons, Higgs particle, and Weak gauge bosons) are set by
the Higgs potential.
\item The light neutrinos are generally believed to get a ``see-saw"
mass of order $v^2/M_R$, where $M_R$ is the scale of lepton-number
violation. But these are negligible for our purposes. 
\item H. Georgi, H. Quinn, and S. Weinberg, {\it Phys. Rev. Lett.}
{\bf 33}, 451 (1974). 
\item See, for example, K.S. Babu and S.M. Barr, {\it Phys. Rev.
Lett.} {\bf 75}, 2088 (1995); G. Anderson, S. Raby, S. Dimopoulos,
and L.J. Hall, {\it Phys. Rev.} {\bf D49}, 3660 (1994); 
Z. Berezhiani, hep-ph/9602325; and references therein.
\item R. Peccei and H. Quinn, {\it Phys. Rev. Lett.} {\bf 38}, 1440 
(1977); S. Weinberg, {\it Phys. Rev. Lett.} {\bf 40}, 223 (1978);
F. Wilczek, {\it Phys. Rev. Lett.} {\bf 40}, 279 (1978).
\item ``The Strong CP Problem: Solutions Without Axions", S.M. Barr,
in {\it CP Violation in Particle Physics and Astrophysics}, ed.
J. Tran Thanh Van (Editions Fronti\`{e}res, 1989), and references
therein.
\item S. Weinberg, {\it Phys. Rev.} {\bf D13}, 974 (1976);
S. Dimopoulos and L. Susskind, {\it Nucl. Phys.} {\bf B155}, 237 (1979).
\item S. Dimopoulos and H. Georgi, {\it Nucl. Phys.} {\bf B193}, 150 
(1981); {\it Phys. Lett.} {\bf 117B}, 287 (1982).
\item V. Agrawal, Ph.D. Thesis, Univ. of Delaware, 1996.
\item R.J. Scherrer and M.S. Turner, {\it Phys. Rev.} {\bf D31},
681 (1985). 
\item V. Kuzmin, V. Rubakov, and M. Shaposhnikov, {\it Phys. Lett.}
{\bf B155}, 36 (1985); {\bf B191}, 171 (1987); P. Arnold and L. McLerran,
{\it Phys. Rev.} {\bf D36}, 581 (1987); {\bf D37}, 1020 (1988).
\item S. Dimopoulos and L. Susskind, {\it Phys. Lett.}
{\bf 81B}, 416 (1979); D. Toussaint, S. Treiman, F. Wilczek, and
A. Zee, {\it Phys. Rev.} {\bf D19}, 1036 (1979); 
S. Weinberg, {\it Phys. Rev. Lett.} {\bf 42}, 850 (1979);
\item For a recent review see A.G. Cohen, D.B. Kaplan, and A.E. Nelson,
{\it Ann. Rev. Nucl. Part. Sci.} {\bf 43}, 27 (1993).
\item J. F. Donoghue, E. Golowich, and B. R. Holstein, {\it Dynamics of the
Standard Model}, (Cambridge University Press, 1992).
\item R. Dashen, {\it Phys. Rev.} {\bf 183}, 1245 (1969);
J. Gasser and H. Leutwyler, {\it Phys. Rep.} {\bf 87},77 (1982);
J. F. Donoghue, in {\bf Building Blocks of Creation - TASI 1993}, 
ed. by S. Raby, (World Scientific,1995) (hep-ph/9403263). 
\item We have explicitly ignored questions 
of composition and taken the number of baryons to be equal to the number 
of fermions. We have also ignored numerical factors that arise from 
geometry (e.g. $4 \pi$) or from more realistic solutions to hydrostatic 
equilibrium which would result in central concentration of the material.
\item With light mesons, and long range nuclear 
reactions, burning may take place at slightly cooler temperatures. This 
is the same issue faced in primordial nucleosynthesis. If screening of 
the nuclear force is effective, then light nuclei exist and stars burn
with $T_N \simgt 1$KeV. If screening is not efficient then the whole discussion 
of stars must be greatly modified.

\end{enumerate}

\section*{Figure Caption}

{\bf Figure 1:} The figure shows a summary of arguments 
that $|\mu^2|<<M_P$ is necessary 
for life to develop. For $\mu^2<0$ large values of $|\mu^2|$ imply large 
values of $v$, and hence larger masses for leptons, quarks, and baryons. 
The increasing difference between the light quark masses, $m_d - m_u 
\propto v/v_0$, implies universes with but a single species
of stable nuclei, which we argue would not allow for chemistry rich 
enough to support life. There is a narrow band where both $p$ and 
$\Delta^{++}$ are stable, but the chemical equivalent of a mix of Hydrogen and 
Helium is plausibly also sterile. For $\mu^2>0$, quark chiral condensates 
lead to $v \propto f_\pi^3/\mu^2$ and quark and lepton masses become very 
small. Light lepton masses imply that biochemical processes cannot occur 
until cosmologically late times, when baryons may have already decayed. 
We show a constraint for a baryon lifetime estimated from exchange of 
intermediate GUT scale ($M_X \approx 10^{16}$ GeV) particles. Even if 
baryons are stable, formation of a biologically acceptable mix of 
elements or the nature of stellar evolution may make development of life 
improbable. What is left is a rather narrow range of $\mu^2 <0$ which 
includes the physical values in our universe.

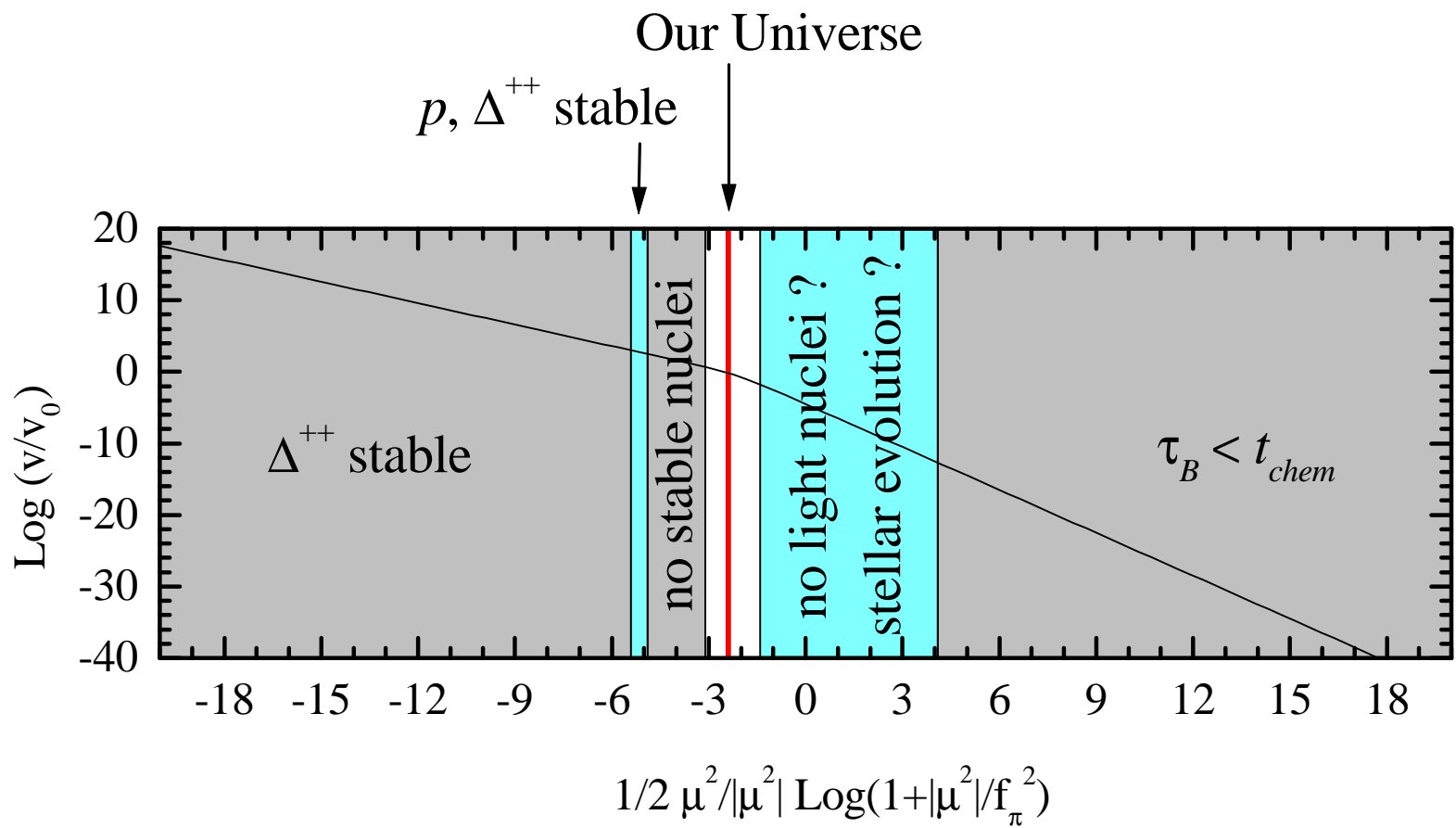
\end{document}